\documentclass[preprint,prc,showpacs,preprintnumbers,superscriptaddress,amsmath,amssymb,floatfix]{revtex4}
\usepackage{graphicx,epsfig,latexsym,amssymb}
\usepackage{multirow,amsmath,array,booktabs}
\usepackage[section]{placeins}
\DeclareGraphicsExtensions{.jpg,.jpeg,.pdf,.png,.mps,.eps,.ps}

\begin{document}
\title{Challenge on the Astrophysical R-process Calculation with Nuclear Mass Models
 \footnote{Partly supported by Major State Basic Research
Developing Program 2007CB815000, the National Natural Science
Foundation of China under Grant Nos. 10435010, 10775004 and
10221003.}}

\author{SUN Baohua$^{1}$\footnote{010-62767013(O), 13810098634}, Meng Jie $^{1,2,3,4}$
 \footnote{mengj@pku.edu.cn, 010-62765620(O)}}
\affiliation{ $^{1}$School of Physics and SK Lab. of Nucl. Phys.
\textsc{\&}
Tech., Peking University, Beijing 100871, China \\
$^{2}$ Department of Physics, University of Stellenbosch,
Stellenbosch, South Africa \\
$^{3}$Institute of Theoretical Physics, Chinese Academy of Sciences,
Beijing, China \\
$^{4}$Center of Theoretical Nuclear Physics, National Laboratory of
Heavy Ion Accelerator, 730000 Lanzhou, China}
 \date{\today}

\begin{abstract}
  Our understanding of the rapid neutron capture nucleosynthesis process
  in universe depends on the reliability of nuclear mass predictions.
  Initiated by the newly developed mass table in the relativistic mean field theory (RMF),
  in this paper the influence of mass models on the $r$-process calculations  is investigated
  assuming the same astrophysical conditions. The different model predictions on the
  so far unreachable nuclei lead to significant deviations in the calculated $r$-process abundances.
\end{abstract}

\pacs{21.10.-k, 21.10.Dr, 21.60.-n, 21.60.Jz, 23.40.-s, 26.30.Hj}
 \maketitle

  The rapid neutron capture ($r$-) process is introduced more than 50 years ago to explain the solar
  abundances not creating from the slow-neutron capture ($s$-) process
  and the proton-capture ($p$-) process~\cite{BBFH}. It is responsible for
  the synthesis of about half of the nuclei beyond the iron group.
  The recent observations~\cite{Cowan-nature06} of extremely
  metal-poor ([Fe/H] $\approx$ -3) $r$-process enriched ([Ba/Eu] $<$ 0)
  stars, show a consistent elemental abundances from Ba to the third $r$-process peak
  with the scaled solar system $r$-process abundance distribution.
  This consistency may indicate that the $r$-process abundance patterns are most possibly produced
  only by a single or at most a few $r$-process events for the heavier elements with $Z \geq 56$,
  i.e., there may be only one or a few $r$-process sites in the early Galaxy.

  However, the exact astrophysical site where the $r$-process proceeds has
  not been unambiguously identified, despite decades of work.
  This research is complicated by the required knowledge of both the astrophysical
  environments and the nuclear properties of very neutron-rich nuclei.
  Previous phenomenological studies indicate that the $r$-process occurs at
  temperatures around T $\sim 10^9$ K and at extreme neutron fluxes
  with neutron number densities $n_n > 10^{20}$ cm$^{-3}$~\cite{Cowan-PR91,Kartz-APJ93,SMG07}.
  Moreover, the $r$-process should be a dynamical process with changing conditions and
  paths~\cite{Kartz-APJ93}.

  On the other hand, the determined astrophysical condition for the
  $r$-process site relies on the extrapolation of theoretical nuclear models
  for the ``terra incognita".
  Among the required nuclear properties, the key one is the nuclear mass,
  from which one can directly determine the one-neutron
  separation energy, shell gap and also the beta-decay energy.
  As discussed in Ref.~\cite{SMG07},
  though various nuclear mass models agree quite well with the known data, they disagree among
  each other towards the very neutron-rich side, where the $r$-process runs along.
  As a result, the required astrophysical condition for
  the $r$-process nucleosynthesis can vary for different model predictions~\cite{SMG07}.

  In order to investigate the impact of nuclear mass models on the $r$-process nucleosynthesis,
  one should distinguish the astrophysical
  uncertainty from the nuclear physics uncertainty.
  In this letter, adopting the newly constructed mass
  table~\cite{RMFBCS} in the relativistic mean field (RMF) theory,
  and assuming the same astrophysical conditions, the impact
  of different mass models on the $r$-process calculation will be
  investigated. Other mass models used include the finite-range droplet model
  (FRDM)~\cite{FRDM}, extended Thomas-Fermi plus Strutinsky integral with
  quenched shell (ETFSI-Q)~\cite{ETFSIQ} and the
  recent Hartree-Fock-Bogolyubov (HFB-13) model~\cite{HFB-13}.

  The RMF approach has made lots of successes in describing
  the nuclear properties far away from the $\beta$-stability line as
  reviewed, for example in Ref.~\cite{Meng06}. The first systematic calculation of the
  nuclear ground state properties including nuclear masses, radii and
  deformations has been done recently for all the nuclei lying between the
  proton drip line and the neutron drip line~\cite{RMFBCS}.
  A good agreement with the available data is found. The detailed analysis shows that
  the predictions of nuclear masses in the RMF are generally underestimated
  with respect to the experimental data.
  However, considering the factor that with less than 10 parameters
  obtained from fitting several doubly magic nuclei,
  the RMF Lagrangian achieves almost
  the same prediction power ($\sim$ 0.65 MeV) for
  one neutron separation energy $S_n$~\cite{AME03} as
  those highly parameterized mass models~\cite{FRDM,ETFSIQ,HFB-13},
  thus it is very interesting to see to what extent the solar $r$-abundances can be reproduced
  using this new table.

\begin{figure*}[th]
   \centering
   \includegraphics[width=15cm]{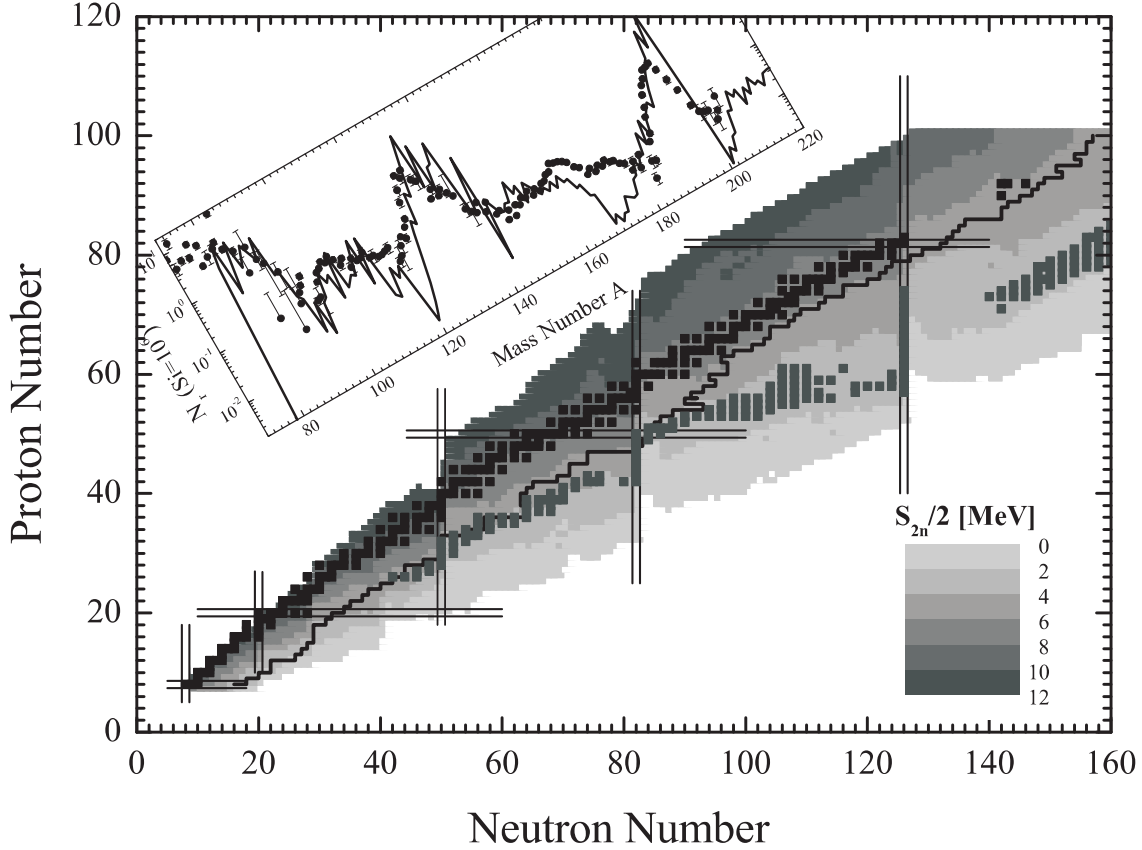}
   \vspace{-10pt} \caption{Features of the $r$-process calculated using
   the new RMF mass table. Black squares denote $\beta$-stable nuclei,
   and magic proton and neutron numbers are indicated by pairs of
   parallel lines. The region in the main graph shows the calculated
   average one neutron separation energy ($S_{2n}/2$). The solid line
   denotes the border of nuclides with known masses in the neutron-rich
   side. The dark grey squares show the $r$-process path when using the
   RMF mass predictions and the FRDM half-lives. The observed and
   calculated solar $r$-process abundance curves are plotted versus the
   mass number A in the inset, whose x-axis is curved slightly to
   follow the $r$-process path.  } \label{fig1}
  \end{figure*}

  We adopt a site-independent $r$-process calculation as in
  Ref.~\cite{SMG07}, where a
  configuration of 16 $r$-process
  components is chosen as a reasonable approximation to the real $r$-process
  site. The seed-nuclei iron are irradiated by neutron
  sources of high and continuous neutron densities $n_n$ ranging
  from $10^{20}$ to $10^{28}$ cm$^{-3}$ over a timescale $\tau$ at
  a high temperature ($T\sim 10^9$ K). The neutron
  captures proceed in (n, $\gamma$)$\leftrightarrow$($\gamma$, n)
  equilibrium, and the abundance flow from one isotopic chain to the
  next is governed by $\beta$-decays. Roughly, the $r$-process runs
  along the contour lines between 2 and 5 MeV of one neutron separation energies as illustrated in
  Fig.~\ref{fig1}. Similar to the classical $s$-process~\cite{kappeler89},
  these different $r$-process components are assumed to satisfy a simply exponential
  formula, i.e.,
  $\omega(n_n)=n_n^a , \tau(n_n)=b\times n_n^c$, where
  $\omega(n_n)$ and $\tau(n_n)$ are the corresponding weighting factor
  and neutron irradiation time for the component with a neutron density $n_n$. The parameters a, b and c
  are determined from the least-square fit to the solar $r$-process abundances.

  The best simulation using the new RMF mass table is presented in
  Fig.~\ref{fig1}. In this simulation, the $\beta$-decay properties are taken from the
  recent calculation~\cite{Moller03}. Furthermore, the available
  experimental results~\cite{AME03,NNDC} have been included.
  The corresponding $r$-process path is indicated by the dark grey
  squares.  The $r$-process abundance distribution,
  in general, can be well reproduced. However, the
  predicted large shell towards the neutron drip line in the RMF model leads to a
  large gap before the second and third abundance peaks. Previous
  investigations~\cite{Kartz-APJ93,pfeiffer-ZPA97} showed that a quenched shell
  could avoid the jump in the $r$-process path and thus result in a better simulation to the
  observation. Nevertheless, it is still unclear whether there is a shell quenching effect
  or to what extent it is towards the neutron drip line, since
  the present experimental results are still in debate.
  One example is the shell gap at the critical waiting point $^{130}$Cd.
  It is firstly suggested to be a quenched shell in Ref.~\cite{dillmann03}.
  However, a recent experiment result~\cite{rising07},
  interpreted by the state-of-the-art nuclear shell-model
  calculations, shows no evidence of shell quenching.

\begin{figure}[th]
\centering \includegraphics[width=8.5cm]{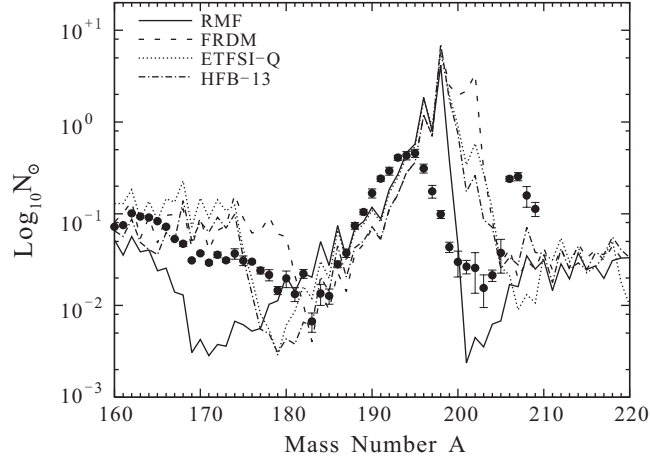} \vspace{-10pt}
\caption{Comparison of observed solar $r$-process abundances (filled
circles) with theoretical abundance after $\beta$-decays calculated
using RMF, FRDM, ETFSI-Q and HFB-13 mass models. The calculated
abundances have been scaled to the solar $r$-process abundance at
$A=130$.} \label{fig2}
\end{figure}

In order to disentangle the nuclear physics uncertainty from the
astrophysical environment uncertainty, we have done the $r$-process
calculations using respectively FRDM~\cite{FRDM},
ETFSI-Q~\cite{ETFSIQ} and HFB-13~\cite{HFB-13} mass inputs while
keeping the same $\beta$-decay properties. All the calculations have
used the same astrophysical condition, i.e., the best case obtained
using the RMF masses. Along with the observation~\cite{kappeler89},
the abundance distributions around the third peak are compared with
the RMF result in Fig.~\ref{fig2}. It shows that the abundances
around the neutron shells $N=126$ strongly depends on the mass
models applied. Before the abundance peak, the RMF simulation shows
a broad dip around $A=170$, while the ETFSI-Q and HFB-13 simulations
have a dip towards a large mass number ($\sim$180). Differences also
exist after the abundance peak.

\begin{figure}[th]
\centering \includegraphics[width=8.5cm]{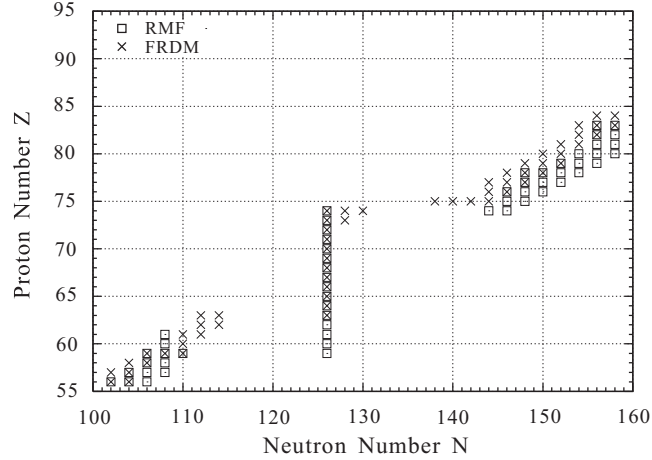} \vspace{-10pt}
\caption{The corresponding $r$-process pathes of Fig.~\ref{fig2} for
the RMF and HFB-13 mass models. Shown are those isotopes with more
than 10\% population of each isotopic chain. For comparison the
stable nuclei are labeled by black squares.} \label{fig3}
\end{figure}

The different deficiencies in reproducing the solar r-process
abundances can be mainly traced back to two aspects. First,
different nuclear mass models predict quite different shell
evolution towards the neutron drip line. The shell gap energy at
$N=126$ in the RMF model is about 2 MeV larger than that in other
models. A stronger shell structure will result in more nuclear
matter accumulated in nuclides with a neutron magic number and less
in nuclides around, which is reflected by dips on both sides of the
abundance peak in the calculated r-process abundance. Second,
different models assign different locations of the shape transition
before the shell closures. The predicted r-process paths pass
through the shape-transition ranges before going to the magic
numbers $N=82$ and 126. The first oblate nuclides for $Z=60$-63 are
in the mass number of 172-177 for the RMF predictions, while in the
HFB-13 model the corresponding mass numbers are 178-179. One neutron
separation energies for these nuclides at the shape-transition point
will deviate from the approximately linear dependence of the mass
number as predicted by a classical liquid droplet mass model, and
eventually affect the r-process path. Together, both aspects result
in a direct jump in the r-process path from $^{169}$Pr to $^{185}$Pr
(15 mass units) in the RMF simulation (see Fig.~\ref{fig3}).
Different from the RMF simulation, abundances calculated in the
HFB-13 simulation are also accumulated in the nuclides with the mass
number 170-177, therefore a better reproduction of the solar
r-process abundances at A$\sim$170, however a worse reproduction at
A$\sim$180. The gap around the $N=126$ shell also exist in other
simulations though varying in magnitude and mass number. In the same
astrophysical environment, the $r$-process path in the RMF
simulation runs about 1-2 mass units towards the neutron drip line
than those of other simulations.

In the present investigation, it is shown that the nature of the
$r$-process is complicated due to the interplay between the nuclear
physics and the astrophysical environment. Since it is still not
accessible to measure most of the nuclei masses along the r-process
path in the near future, further theoretical development aiming at
the description of the know and unknown masses simultaneously is
badly needed. If the astrophysical conditions for the $r$-process
are identified precisely, this may serve as a constraint for the
nuclear mass models. Equivalently, if nuclear masses are known in a
good precision, it can be used to constraint the potential site for
the $r$-process as well.

In the present letter, we have shown that the $r$-process
calculation is quit sensitive to the nuclear mass inputs. Precise
mass measurement of neutron-rich nuclides with an accuracy less than
100 keV are needed to decisively determine the shell evolution at
$N=82$ and 126 towards the neutron drip line, as well as the
locations of the shape transition before these shell closures. These
experimental results will offer a primary constraint to the existing
mass models and a strong motivation for further exploration of
theoretical mass models, and furthermore, a better understanding the
nature of the r-process nucleosynthesis.. Meanwhile, as there is no
clear evidence to accept or reject any mass model mentioned above,
it is necessary to take different mass models into account in the
$r$-process study.


\end{document}